\documentclass[11pt,twoside]{article}

\usepackage{asp2006}
\usepackage{epsf}
\usepackage{lscape}

\begin{document}
\title{Forty Years of Research on Isolated Galaxies}   %%% Fill in title
\author{J. Sulentic}
%%% Fill in author names
\affil{$^1$Instituto de Astrof\'isica de Andaluc\'ia}    %%% Fill in author affiliations

\begin{abstract} %%% Abstract to run on from here.
Isolated galaxies have not been a hot topic over the past four decades. This is partly due to uncertainties about their
existence. Are there galaxies isolated enough to be interesting? Do they exist in sufficient numbers to be
statistically useful? Most attempts to compile isolated galaxy lists were marginally successful--too small number and not
very isolated galaxies. If really isolated galaxies do exist then their value becomes obvious in a Universe where
effects of interactions and environment (i.e. nurture) are important. They provide a means for better quantifying effects
of nurture. The Catalog of Isolated Galaxies (CIG) compiled by Valentina Karachentseva appeared near the beginning of the review
period. It becomes the focus of this review because of its obvious strengths and because the AMIGA project has increased
its utility through a refinement (a vetted CIG). It contains almost 1000 galaxies with nearest neighbor crossing times of 1-3Gyr.
It is large enough to serve as a zero-point or control sample. The galaxies in the CIG (and the distribution of galaxy types)
may be significantly different than those in even slightly richer environments. The AMIGA-CIG, and future iterations, may be
able to tell us something about galaxy formation. It may also allow us to better define intrinsic (natural) correlations
like e.g. Fisher-Tully and FIR-OPTICAL. Correlations can be better
defined when the dispersion added by external stimuli (nurture) is minimized or removed.

\end{abstract}

%%% MAIN BODY OF TEXT GOES HERE. CONSULT "INSTRUCTIONS FOR AUTHORS USING
%%% LATEX2E MARKUP", SECTIONS 2.3-2.6 FOR HELP WITH EQUATIONS, FIGURES,
%%% AND TABLES.

\section{Introduction}

Studies of isolated galaxies can be said to begin at about then
same time it was realized that galaxy `nurture' was important
(e.g. Sulentic 1976; but see also Tovmassian 1966). I remember
well the criticism for suggesting in my thesis that radio emission
from interacting galaxies was enhanced. Of course within a decade
the role of nurture on galaxy structure and evolution was
established (Larson \& Tinsley 1978; Stocke 1978; Lonsdale et al.
1984; Cutri \& McAlary 1985). There remains some confusion about
the nurture mechanism(s) which stems in part from the fact that
there are two definitions of nurture: 1) galaxies can be
influenced by the local environmental surface density (e.g.
clusters to voids) and 2) galaxies in any environment can be
influenced by one or more nearby companions (pairs and compact
groups). The morphology-density relation (Postman \& Geller 1984)
formalizes evidence for definition 1 while numerous theoretical
and observational studies (e.g. Keel 2009) confirm
definition 2. Definition 1 deals largely with the frequency of
occurrence of
early-type galaxies as a function of environmental
density while Definition 2 deals largely with late-types whose
non-stellar components manifest most clearly the effects of
"one-on-one" nurture at all environmental densities (interacting
pairs are found in both clusters and voids). It has been argued
that all early-types are the product of late-type merging and
harassment (e.g. Barnes 1998; Schweizer 2000). We argue here that
the evidence (the existence of isolated early-types) does not
support this view.

Once it is agreed  that galaxy structure and evolution are influenced by
environment then the value of a carefully selected isolated sample becomes
obvious. Some researchers argue that truly isolated galaxies do not exist but
this is often simply a matter of semantics. It is now generally accepted that
there is no field galaxy population (Huchra \& Thuan 1977, Einasto 1990). Rather
than speaking of a specific class of isolated galaxy it is perhaps more useful
to speak of identifying the most isolated galaxies that exist. Many attempts to
identify these galaxies yielded samples too small for statistical applications.
One sample, the Catalog of Isolated Galaxies (CIG: Karachentseva 1973), is large
enough (n$\sim$1050) to serve as a useful statistical control for studies
seeking to quantify the effects of nurture. In fact the CIG sample serves as a
counterpoint to the Catalog of Isolated Pairs of Galaxies published one year
earlier (CPG: Karachentsev 1972). I became aware of these catalogs in 1974,
too late to use them in my thesis (instead see Stocke 1978). Soviet publications
were not available in every university library in the US and there was no internet
in the early 70's.

\section{Lessons from the 70's, 80's and 90's}

Publication of the visually compiled CIG corresponds in time to the first attempts at
automated catalog compilation. Surprisingly no automated catalog appeared over the next
30 years that could replace the CIG in quality or completeness. Looking back we appreciate
that algorithms cannot select well isolated systems whether singlet, pair, triplet or compact
group. There are two main reasons for this limitation: 1) source catalogs used by an algorithm
are flux limited and many close companions to candidate isolated galaxies  will lie below that
limit and 2) close bright companions are sometimes not listed separately in source catalogs.
Algorithms work well in terms of (above) isolation definition 1 but the product catalogs
include many violations of isolation definition 2 (e.g. Turner \& Gott 1975, Vettolani et al. 1986).
Visual verification of product catalogs is always necessary and this is why the CIG (and CPG)
remain valuable resources to this day. The CPG satisfies definition 1 but completely fails
definition 2 while the CIG satisfies both.

The CIG has a large number of strengths: WELL DEFINED: visual application of a precise isolation criterion;
SIZE: 800-1050 galaxies depending on how one defines isolation -- permitting exploration of the luminosity-morphology
domain of isolated galaxies; REDSHIFT: almost complete coverage; ISOLATION: nearest neighbor crossing
times t$_c$= 1-3Gyr; DEPTH/COMPLETENESS: 80-90\% complete to V$_r$=10000km/s (for detailed analyses of the CIG
prior to the AMIGA project see Adams et al. 1980, Karachentseva 1980, Xu \& Sulentic 1991). Pre-AMIGA radio line/continuum,
optical and infrared studies of various CIG subsets can be found by inserting ``isolated galaxies'' in the title
query panel after entering adsabs.harvard.edu  and proceeding to the ``astronomy and astrophysics search page. Such a
query will also lead one to other lists of isolated galaxies. Most, including the CIG, contain obvious examples of Definition 2
violations. Visually compiled lists are only as good as the images from which they are compiled--in $\sim$ 1970 the best
image resource was the Palomar Sky Survey (POSS). It is interesting to note the small number of CIG galaxies included in
(northern sky overlap) most of the other lists (e.g. Xanthopoulos \& de Robertis 1991, 1/8 Seyferts; Morgan et al.
1998, 3/12 spirals; Aguerri 1999, 6/16 barred; Pisano \& Wilcots 1999, 0/6
extremely isolated; Marquez \& Moles 1996, 1999, spirals 4/22; Colbert et al.
2001, 1/18 early-types; Kornreich et al. 2001, 1/9 SA type; Pisano et al.
2002, 4/22 extremely isolated; Madore et al. 2004, 1/12 ellipticals
from Bothun et al. 1977; Reda et al. 2004, 2/13 isolated early-type).
The small overlap  does not in the majority of cases, reflect isolated galaxies
overlooked by the CIG but rather the more stringent isolation criteria used to compile the CIG.

Isolated galaxies have never been a hot topic yet, if real, a  proper list would represent a
valuable resource. One of the few large pre-AMIGA series of papers dealing with reasonably isolated
galaxies involves samples of from 22-203 isolated spiral galaxies (Varela et al. 2004,
Marquez et al. 2004 and references back to 1996).  While making no claims about
sample completeness the authors  present interesting comparisons with less-isolated and more-active
spirals. They found that isolated spirals tend towards later-types, symmetric morphologies. They may be bluer,
smaller and less luminous (but see later). This work also finds tighter Kormendy and Fisher-Tully relations
for the isolated galaxies if, as might be expected, nurture adds dispersion to most galaxy measures.
Consistent with that assumption they also find a narrower range of disk scale length and effective
surface brightness ($\mu_e$) for more isolated galaxies.

\section{The AMIGA project}

We have summarized the advantages of the CIG but there are also problems, principally that it was compiled
from a visual search of the original Palomar Sky Survey (POSS). Visual compilation is less a problem than the low
resolution and non-linearity of the POSS data. This means that the originally assigned galaxy types are OK but
spiral galaxy bulges appear larger on low resolution images leading to an overestimate of the early-type spiral fraction.
Compact spirals can easily be misclassified as E or S0 type. Compact companion galaxies can be overlooked.  The
strengths of the catalog motivated us to undertake a systematic refinement or ``vetting'' of the
CIG under the AMIGA ({\bf A}nalysis of the Interstellar {\bf M}edium of {\bf I}solated {\bf GA}laxies
{\tt http://amiga.iaa.es}) project. We expected to reject a significant number of CIG but were confident
that a sample large enough for statistical studies would remain.  We have focused so far on galaxies with recession
velocities V$_r$ greater than 1500km/s. The largely local and largely dwarf galaxy population within 1500km/s
deserves separate consideration. Excluding the local objects and non-isolated galaxies yields a final vetted sample
of 719 galaxies. We summarize the steps in the AMIGA project  that have lead to a refined CIG sample. We also summarize
the initial multi wavelength studies of the sample.

\begin{itemize}

\item OLF derivation:  (Verdes-Montenegro et al. 2005) Application of a V/V$_m$ test suggests that
the vetted CIG is $\sim$90\% complete to B$\sim$15.0. The histogram of heliocentric velocities shows
peaks identifiable with local large scale structure components (within 10-15$\times$10$^4$ km/s)
(see also Haynes \& Giovanelli 1983). However underlying the peaks we find that about half
of the sample show a quasi-homogeneous distribution. While a galaxy "field" may not exist we identify
a very field-like galaxy population. Comparison of the CIG OLF with other derivations indicates that the
best fit Schechter M*= -20.1  parameter is as low or lower than previous estimates (except 2dFGRS void).

\item Morphology Refinement: (Sulentic et al. 2006) Morphological vetting made use of the POSS2 (using SDSS
CCD estimates for 215 galaxies to evaluate accuracy) for the n=1018 galaxies remaining after the optical
OLF study. Fully 2/3 of the CIG sample are found to be late-type (i.e. small bulge) Sb-Sc spirals. 15\% show
later spiral types yielding a late-type fraction of $\sim$86\%. N=32 galaxies were identified as
interacting (Definition 2 nurture) while an additional 154 were flagged as possibly interacting systems.
Perhaps most interesting are the 14\% of the sample classified as early-type and distributed approximately equally
between elliptical and lenticular galaxies. We find a wide range of luminosities (M$_B$=-17 to -23) among the
Sb-Sc majority with M*= -20.2. Perhaps most surprisingly the E+S0 galaxies show a similar M* with a range
M$_B$=  -16 to -21 and only two galaxies near -22. These are perhaps the best candidates for a primordial
early-type population since they are low luminosity systems and are found in environments where merging and
harassments are improbable (see also Sulentic \& Rabaca 1994). We also compare our results with those for several
large surveys including 2dFGRS, SDSS, NOG and SSRS2 (see paper for detailed references and discussion). Such comparisons
are difficult because different definitions of isolation (i.e. voids) and different selection techniques (Definition 1)
were employed. It is interesting to note that the galaxy population in the lowest density bin of 2dFGRS is similar in
size to the CIG.

\item  MIR/FIR Statistics: (Lisenfeld et al. 2007) Mid- and Far-infrared emission from galaxies has proven to be perhaps the most
sensitive diagnostic of nurture. The wide dispersion of MIR/FIR values in isolated and interacting samples makes it difficult to
improve the accuracy of nurture quantification (Xu \& Sulentic 1991). If the vetted CIG is almost nurture free then it offers
the possibility to serve as the optimal control sample in such studies. Indeed mean FIR luminosity (logL$_{FIR}$=9.5L$_{\odot}$)
and FIR - Optical flux ratio (log R= 0.44)  for the Sb-Sc majority show among the lowest mean values of any galaxy sample
yet studied. The absence of nurture contamination also means that the dispersion in measured values is lower--a desideratum for a good control sample. An internal comparison was made by comparing mean values for vetted CIG and for 14 CIG flagged as interacting during the morphology phase. The latter galaxies showed much higher mean values (9.9 and -0.06  respectively) illustrating the effect of including nurture contaminated galaxies in a control sample.

\item Isolation Refinement: (Verley et al. 2007ab) The vetted CIG is shown to be a sample where nearest neighbor crossing times are t$_c$$\sim$1-3Gyr.
See contribution by Simon Verley in this volume.

\item Radio Continuum Emission: (Leon et al. 2008) The radio detection fraction and derived radio luminosities
for detections reveal a very radio-quiet sample. Most detections show radio luminosities in the range
L$_{1.4GHz}$=20-22 W Hz$^{-1}$ with only three sources showing a radio/optical flux ratio R$\geq$100 (i.e. formally  a radio-galaxy
according to Kellerman et al. (1989) (most detections lie between R=1-10). Comparison of detections and
fluxes in the NVSS and FIRST surveys suggests that the radio emission arises in most cases from disk emission driven by star formation.

\item AGN in Isolated Galaxies: (Sabater et al. 2008) Using various techniques to identify active nuclei yields a list of 89 candidates
which implies the lowest AGN fraction that we are aware of. See contribution by Pepe Sabater in this proceeding. See also
Deborah Dultzin's review for discussion of the AGN--nurture question.

\item 21cm HI Profiles: (Espada et al. in preparation) The Arecibo HI survey (Haynes \& Giovanelli 1980, 1983, 1984; Hewitt et al. 1983)
explored HI properties of a large CIG subsample and found that HI mass depends more on galaxy
diameter than morphological type. In fact examination of their measures reveals no difference in HI measures
between the Sb, Sbc and Sc galaxies which dominate the CIG. The AMIGA HI survey incorporates a much larger sample
of old and new data. We find a very low fraction of asymmetric profiles as expected for an isolated sample. For the first time we
are able to isolate the intrinsic (non-nurture) asymmetry distribution. See contribution by Dani Espada in this proceeding.

\item Studies of Subsamples: More recent photometric/structural analysis using SDSS imagery have focussed on 100 of
the Sb-c galaxies in the vetted CIG (M$_i$= -19 to -23). Both
surface photometric (Durbala et al. 2008) and Fourier (Durbala et al. 2009) analyses were
performed on this `prototypical' isolated
sample. Most bulge/total flux ratios were found to lie B/T$\leq$ 0.2 with Sersic
indices generally n$_{BULGE}$$\leq$ 2.5 which is consistent with the idea that most isolated
spirals contain a pseudo bulge. A correlation between B/T and  n$_{BULGE}$  is seen only for
galaxies visually classified Sb (largest bulge/disk ratios) suggesting that some of these
may involve nurtured bulges. Overall more than 90$\%$ of the sample may contain
pseudo bulges and subjective type Sb may represent a boundary between largely classical and largely
pseudo bulge populations. Bar-spiral decomposition was accomplished using Fourier processing of
the images. Comparison with the OSUBGS sample (Eskridge et al. 2002) indicates that the CIG contains
many more large bars (l$_{bar}$$\geq$ 3kpc). The analysis reveals a 4$\times$ higher fraction of
spirals dominated by m=2 \& 3 multiplicity compared to the CSRG sample (Buta 1995). The first step
in evaluating H$\alpha$ emission properties of AMIGA-CIG galaxies involves a study of the 45
largest and least inclined spirals (Verley et al. 2007c) where we estimate torques between the gas and
stellar mass components.

\end{itemize}

\section{Future Work}

This review has unavoidably emphasized results from the AMIGA project and passed over detailed studies of individual isolated galaxies.
The size of the sample and the strength of the isolation criterion make it the best place to begin statistical studies
involving the low density tail of the two point correlation function and/or morphology density relation. Analysis of automated
surveys is fraught with peril until they can be properly vetted with good imagery--SDSS offers great promise).
Review of the literature shows that isolation means different things to different people. The label
itself is a source of debate. Do isolated galaxies really exist?  Have they always been isolated? Rather
than using the sobriquet ``isolated'' we suggest that the AMIGA vetted CIG is simply the best compilation
of the most isolated galaxies in the local Universe. Nearest neighbor crossing times of 1-3Gyr are long enough
to be interesting. The fact that we find multi wavelength differences with other samples (even those labeled isolated) argues that
these galaxies have spent most of their lives in relative isolation. The low luminosity early-type fraction and the large fraction of small
bulge spirals cannot have experienced much nurture in their lives. This is the closest we are likely to
come to identifying a population of galaxies whose properties are dominated by nature rather than nurture.

Other isolated samples tabulated earlier show little overlap with the CIG. Most of them report little
difference when compared to samples in richer environments. These samples appear to contain more luminous early types
and spirals with large bulges than found in the CIG. This leads to the possibility that galaxy properties change
significantly below a certain density threshold--dare we say ``field-like''?. There are two ways to explore this
possibility: 1) the late-type (Sb-c) spiral majority and 2) the early-type (E+S0) population. The former
prototypical population requires
testing of the hypothesis that it is dominated by pseudo bulges. This would obviously favor bottom-up spiral galaxy formation.
Measures of bulge colors and kinematics can provide direct tests of the pseudo-bulge hypothesis. In the case of the low
luminosity early-types we need to confirm that they are indeed E and S0 systems. The relative fractions of these
two types needs to be better determined and, finally, spectroscopy to infer their star formation histories.
Initial work (Marcum et al. 2004) for nine CIG ellipticals yields mixed results with some showing typical red
colors and others much bluer than normal. I see a couple nice thesis projects here!

\acknowledgements We wish to acknowledge Junta de Andalucia contract
                  P08-FQM-4205-PEX and AYA Project AYA2008-06181-CO2.

\end{document}